\documentclass[aps,prb,showpacs,bibnotes,groupedaddress,showkeys,reprint]{revtex4-1}

\usepackage{graphicx}
\usepackage{amsmath}
\usepackage{bm}
\usepackage{dcolumn}
\usepackage{amsfonts}
\usepackage{amssymb}
\usepackage[table]{xcolor}
\usepackage{tabularx}
\renewcommand{\tabcolsep}{0.058cm}

\begin{document}
\title{Spin-dependent Seebeck coefficients of Ni$_{80}$Fe$_{20}$ and Co in nanopillar spin valves}
\author{F. K. Dejene}
\email[Electronic mail: ]{\textsl{ f.k.dejene@rug.nl}}
\author{J. Flipse}
\author{B. J. van Wees}
\affiliation{\textit{Physics of Nanodevices, Zernike Institute for Advanced Materials, University of Groningen, Groningen, The Netherlands}}

\date{\today}

\begin{abstract}
We have experimentally determined the spin-dependent Seebeck coefficient of permalloy (Ni$_{80}$Fe$_{20}$) and cobalt (Co) using nanopillar spin valve devices. The devices were specifically designed to completely separate heat related effects from charge related effects. A pure heat current through the nanopillar spin valve, a stack of two ferromagnetic layers (F) separated by a non-magnetic layer (N), leads to a thermovoltage proportional to the spin-dependent Seebeck coefficient $S_{S}$$=$$S_{\uparrow}$$-$$S_{\downarrow}$ of the ferromagnet, where $S_{\uparrow}$ and $S_{\downarrow}$ are the Seebeck coefficient for spin-up and spin-down electrons. By using a three-dimensional finite-element model (3D-FEM) based on spin-dependent thermoelectric theory, whose input material parameters were measured in separate devices, we were able to accurately determine a spin-dependent Seebeck coefficient of $-$1.8 $\mu$V K$^{-1}$ and $-$4.5 $\mu$V K$^{-1}$ for cobalt and permalloy, respectively corresponding to a Seebeck coefficient polarization $P_{S}$$=$$S_{S}/S_{F}$ of 0.08 and 0.25, where $S_{F}$ is the Seebeck coefficient of the ferromagnet. The results are in agreement with earlier theoretical work in Co/Cu multilayers and spin-dependent Seebeck and spin-dependent Peltier measurements in Ni$_{80}$Fe$_{20}$/Cu spin valve structures.
\end{abstract}


\maketitle

\section{\label{sec:level1}Introduction}

The interplay between spin and heat transport in magnetic structures is studied in the emerging field called spin caloritronics.\cite{bauer_spin_2010,bauer_spin_2012} This subfield of spintronics has recently gained a lot of interest leading to notable experimental\cite{uchida_observation_2008,Lebreton_thermal_2011,walter_seebeck_2011,slachter_thermally_2010,flipse_direct_2012,yu_evidence_2010} and theoretical studies.\cite{hatami_thermoelectric_2009,scharf_theory_2012} At the heart of spin caloritronics lie the spin-dependent Seebeck and related spin-dependent Peltier effect. The spin-dependent Seebeck effect describes thermally driven spin injection from a ferromagnet (F) into a non-magnetic (N) material when the F/N interface is subjected to a temperature gradient. This effect is governed by the difference in the Seebeck coefficients of spin-up electrons $S_{\uparrow}$ and spin-down electrons $S_{\downarrow}$. Slachter \textit{et al.}\cite{slachter_thermally_2010} demonstrated the spin-dependent Seebeck effect in Ni$_{80}$Fe$_{20}$/Cu lateral spin valve devices from which a spin-dependent Seebeck coefficient $S_{S}$$=$$S_{\uparrow}$$-$$S_{\downarrow}$ of $-$3.8 $\mu$V K$^{-1}$ was extracted using a 3D-FEM. Here it is important to point out the fundamental difference between the spin-dependent Seebeck and the so called `spin Seebeck effect'.\cite{uchida_spin_2008} Whereas the spin-dependent Seebeck effect is purely electronic in nature, the latter is now understood to originate from collective effects involving non-equilibrium thermally induced spin pumping due to temperature differences between, for example, conductions electrons and magnons.\cite{xiao_theory_2010, bauer_spin_2012}

The spin-dependent Peltier effect, which is the reciprocal of the spin-dependent Seebeck effect, describes heating/cooling of a F/N interface by a spin current. More recently, Flipse \textit{et al.}\cite{flipse_direct_2012} demonstrated the spin-dependent Peltier effect in Ni$_{80}$Fe$_{20}$/Cu/Ni$_{80}$Fe$_{20}$ nanopillar spin valve devices from which a spin-dependent Peltier coefficient $\Pi_{S}$ of $-$1.1 mV was obtained. The spin-dependent Seebeck and Peltier coefficient reported in Refs.~\onlinecite{slachter_thermally_2010} and \onlinecite{flipse_direct_2012} follow the Thomson$-$Onsager relation $\Pi_{S}$$=$$S_{S}T_{o}$, where $T_{o}$ is the temperature.

Although the concept of the spin-dependency of the Seebeck coefficient was first discussed by Campbell \textit{et al.}\cite{campbell_transport_1982} and later used to explain large magnetothermoelectric powers in multilayers of Co/Cu \cite{shi_giant_1993,baily_magnetothermopower_2000,gravier_spin-dependent_2004}, reports on the Seebeck coefficient polarization $P_{S}$=$S_{S}/S_{F}$ are relatively scarce. For Ni$_{80}$Fe$_{20}$, a $P_{S}$ of 0.20 has been reported from spin-dependent Seebeck\cite{slachter_thermally_2010} and spin-dependent Peltier\cite{flipse_direct_2012} measurements. In case of Co, effective $P_{S}$ values ranging from 0.18\cite{gravier_thermodynamic_2006,gravier_spin-dependent_2004} to 0.42\cite{shi_giant_1993,cadeville_thermoelectric_1971} were reported from thermopower measurements in Co/Cu multilayers and diluted Co alloys, respectively. To quantify the size of spin caloritronic effects, one needs to accurately determine spin-dependent thermoelectric coefficients. In this paper, therefore, we provide absolute values of the spin-dependent Seebeck coefficient and its polarization for cobalt and permalloy from spin-dependent Seebeck measurements in F/N/F pillar spin valve devices.

The objectives of this paper are therefore twofold. First, it describes the spin-dependent Seebeck effect in specifically designed nanopillar spin valve devices. Secondly, it presents an accurate determination of the spin-dependent Seebeck coefficients for Ni$_{80}$Fe$_{20}$ and Co using a 3D-FEM. To that end, the electrical conductivity and Seebeck coefficient of all materials were measured in separate devices. The thermal conductivity of the thin metallic films was obtained from the measured electrical conductivity by using the Wiedemann-Franz law.\cite{bakker_thermal_2012} Thermal conductivity of insulating layers was determined from heat transport measurements across metal/insulator/metal structures.

This paper is organized as follows. In sec.~\ref{sec:level2}, we present general spin-dependent thermoelectrics in the framework of the two spin channel model and particularly explain thermally driven spin injection in symmetric F/N/F nanopillar devices. We also discuss the improvements to the 3D-FEM in terms of separately measuring the inputs material parameters. Sec. \ref{sec:level3} presents details of the device fabrication and measurement schemes used in this study. Here we also explain how we achieve a temperature gradient over the F/N/F stack and present the two types of measurements that were performed to fully characterize the devices. Sec. \ref{sec:level4} presents the results of the electrical and thermal spin injection experiments and discuss how the polarization of the conductivity and of the Seebeck coefficient were extracted using the 3D-FEM. Finally, Sec. \ref{sec:level5} presents the conclusions.
\section{\label{sec:level2}Spin-dependent Seebeck effect in F/N/F pillar spin valve}
\begin{figure}[t]
\includegraphics[height=3.6cm]{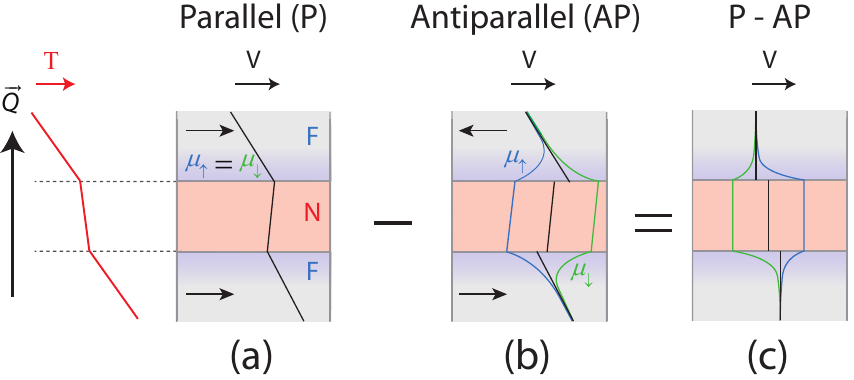}
\caption{\label{fig:fig1}(Color online) Spin electrochemical potentials $\mu_{\uparrow}$ (blue) and $\mu_{\downarrow}$ (green) in a F/N/F stack subjected to a temperature gradient in the case when the the magnetization are aligned (a) parallel and (b) antiparallel. (c) shows the difference between (a) and (b). The heat current and temperature profile are also shown to the left of (a).}
\end{figure}
In metallic ferromagnets, charge, spin and heat transport can be described by two parallel spin channels, one for spin-up ($\uparrow$) and another for spin-down ($\downarrow$) electrons, with each spin channel having its own conductivity $\sigma_{\uparrow,\downarrow}$ and Seebeck coefficient $S_{\uparrow,\downarrow}$.\cite{macdonald_thermoelectricity_1962,campbell_transport_1982} The charge and heat current in each spin channel are related to their respective potential gradient $\vec{\nabla}\mu_{\uparrow,\downarrow}$ and temperature gradient $\vec{\nabla}T$ as~\cite{slachter_thermally_2010}
\begin{equation}
\left( \begin{array}{c} \vec{J_{\uparrow}} \\ \vec{J_{\downarrow}} \\ \vec{Q} \end{array} \right) =
-\left( \begin{array}{ccc} \sigma_{\uparrow} & 0 & \sigma_{\uparrow}S_{\uparrow} \\ 0 & \sigma_{\downarrow} & \sigma_{\downarrow}S_{\downarrow} \\ \sigma_{\uparrow}\Pi_{\uparrow} & \sigma_{\downarrow}\Pi_{\downarrow} & k \end{array} \right)\left( \begin{array}{c} \vec{\nabla}\mu_{\uparrow}/e \\ \vec{\nabla}\mu_{\downarrow}/e \\ \vec{\nabla} T \end{array} \right),
\label{Eq:3current}
\end{equation}
where $\Pi_{\uparrow,\downarrow}$ and $\mu_{\uparrow,\downarrow}$ are the Peltier coefficient and electrochemical potential for spin-up and spin-down electrons and $\kappa$ is the thermal conductivity. Equation \eqref{Eq:3current} is the basis for our 3D-FEM, which was previously used to describe spintronic and spin caloritronic phenomena. A detailed procedure for the modeling can be found in Ref.~\onlinecite{slachter_modeling_2011}. By separately measuring the modeling parameters for each material in dedicated devices\cite{bakker_thermal_2012}, good agreement between the model and the measurement was obtained allowing us to accurately determine the spin-dependent Seebeck coefficients by using the measured electrical and thermal spin signals.
\begin{figure}
\includegraphics[height=3.6cm]{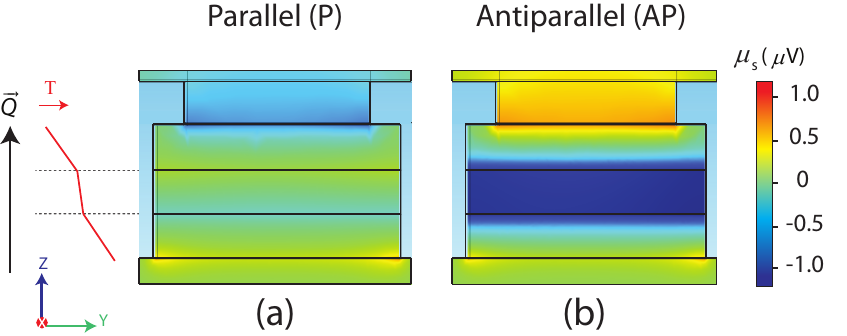}
\caption{\label{fig:fig2}(Color online) ZY-plane cross-section plot obtained from the 3D-FEM of the spin accumulation $\mu_{s}$$=$$\mu_{\uparrow}$$-$$\mu_{\downarrow}$ through the middle of a Ni$_{80}$Fe$_{20}$/Cu/Ni$_{80}$Fe$_{20}$ nanopillar for a temperature change $\Delta T$$=$$7$ K across the the stack for the (a) parallel and (b) antiparallel configurations. The spin accumulation of --1 $\mu$V in (b) is significantly larger than in (a).}
\end{figure}

In the following, we describe the spin-dependent Seebeck effect in a symmetric F/N/F pillar stack with equal layer thicknesses ($t$=15 nm) comparable to the spin relaxation length in the ferromagnet ($\lambda_F$) but much smaller than in the non-magnetic layer ($\lambda_N$=300 nm). In a ferromagnet, owing to the difference in the spin-dependent Seebeck coefficients $S_{\uparrow}$$\neq$$S_{\downarrow}$, a temperature gradient $\nabla T$ across a F/N interface drives a spin current $J_{\uparrow}$$-$$J_{\downarrow}$ from the F-- into the N--region\cite{slachter_thermally_2010} thereby creating a non-equilibrium spin-accumulation $\mu_{s}$$=$$\mu_{\uparrow}$$-$$\mu_{\downarrow}$, which is proportional to the spin-dependent Seebeck coefficient $S_{S}$$=$$S_{\uparrow}$$-$$S_{\downarrow}$ of the ferromagnet. Here, we define spin-up electrons as the spins with the higher conductivity, which in case of both permalloy and cobalt are the majority spins. For a F/N/F pillar stack in a temperature gradient, thermal spin injection at the the two F/N interfaces results in a spin accumulation in the N--region that is a function of the relative alignment of the magnetization of the ferromagnets.

In the parallel ($\uparrow \uparrow$) configuration (Fig.~\ref{fig:fig1}(a)), spins are injected at the first interface while being extracted at the second resulting in a flow of constant spin current across the whole stack. This constant spin current flow dictates that there is negligible spin accumulation at the two F/N interfaces, that is, the individual spin chemical potentials $\mu_{\uparrow}$ and $\mu_{\downarrow}$ are equal. In the antiparallel configuration ($\uparrow \downarrow$), however, spins of similar kind are injected from both interfaces into the N--region. In such configuration, the spin current in the bulk of the ferromagnets is opposite to each other giving rise to a large spin accumulation in the N--region. This large spin accumulation results in the splitting of the spin electrochemical potentials (see Fig.~\ref{fig:fig1}(b)). A cross-sectional plot of the spin accumulation $\mu_{s}$  obtained from the three-dimensional FEM (shown in Fig.~\ref{fig:fig2}) demonstrates the significant difference in the size of the spin accumulation for the two different configurations.

An expression for $\mu_{s}$, based on one--dimensional spin-diffusion equation, in the limit $t$ $\gg$ $\lambda_{F}$,$\lambda_{N}$, can be found elsewhere.\cite{scharf_theory_2012,slachter_thermally_2010} Here we extend this limit to devices with thicknesses $t$ comparable to $\lambda_{F}$ and $\lambda_{N}$ and find the expression given in Eq.~\ref{eq:appendix3} of the appendix, which is similar to the expression in Ref.~\onlinecite{slachter_thermally_2010} except for the resistance mismatch factor. The interfacial spin thermoelectric voltage drop $\Delta \mu$$=$$P_{\sigma}\mu_{s}$, which is different for the two configurations, can then be expressed as a function of the spin accumulation at the two F/N interfaces. In an experiment, one measures this open-circuit thermovoltage as a function of an external magnetic field. The spin valve signal $V_{SV}$$=$($\Delta \mu^{\uparrow \uparrow}$$-$$\Delta \mu^{\uparrow \downarrow}$)/e is thus given by:
\begin{equation}
V_{SV}=-2\lambda_{F} S_S\nabla T P_{\sigma}R_\text{mismatch},
\label{eq:spinaccumuulation1}
\end{equation}
where $\nabla T$ is the temperature gradient in the F-region. The term $R_\text{mismatch}$ denotes the resistance mismatch factor for a symmetric spin valve given by:
\begin{equation}
R_\text{mismatch}=\frac{\cosh(\frac{t}{\lambda _F})-\text{exp}(-\frac{2 t}{\lambda _F})}{ \frac{R_F}{R_N}\cosh(\frac{t}{\lambda _F})\tanh(\frac{t}{2 \lambda _N})+\sinh(\frac{t}{\lambda _F})},
\label{eq:mismatch}
\end{equation}
where $R_{F}$$=$$\frac{\lambda_{F}}{(1-P_{\sigma}^{2})\sigma_{F}}$ and $R_{N}$$=$$\frac{\lambda_{N}}{\sigma_{N}}$ are the spin-resistances of the ferromagnet and the normal metal. In the limit $t$ $\gg$$\lambda_{F}$,$\lambda_{N}$, $R_\text{mismatch}$ reduces to the single F/N interface result which is often close to one. Note, however, that in the analysis we use the numerical results from the three-dimensional finite element modeling based on Eq.~\ref{Eq:3current} to extract $P_{\sigma}$ and $P_{S}$.
\section{\label{sec:level3}Experiments}
\begin{figure}[h]
\includegraphics[width=8.6cm]{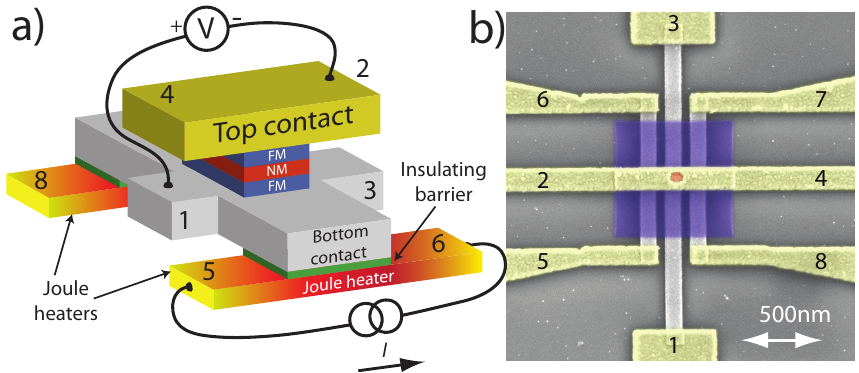}
\caption{\label{fig:fig3}(Color online) (a) Schematic representation of the measured device showing a F/N/F stack sandwiched between a Au-top contact (yellow) and Pt-bottom contact (grey). Platinum Joule heaters, which are electrically isolated from the bottom contact by an AlO$_{x}$ barrier (green), are used to heat the bottom of the nanopillar. Homogeneous heating is achieved by two Pt Joule heaters on either side of the nanopillar. (b) Colored scanning electron microscope image of the measured device. Cross-linked PMMA matrix (blue) surrounding the pillar (red) is used to isolate the bottom contact from the top contact}
\end{figure}

The nanopillar spin valve devices were prepared in one optical lithography step followed by nine electron-beam lithography (EBL) steps. Materials were deposited by e-beam evaporation at a base pressure of $2\times 10^{-6}$ Torr on a thermally oxidized Si substrate with a 300-nm-thick oxide layer. Fig.~\ref{fig:fig3}(a) and Fig.~\ref{fig:fig3}(b) show a schematic and scanning electron microscope image of the measured device. The device consists of a F/N/F stack sandwiched between a bottom and top contact. The experimental methods and device fabrications are similar to the ones reported in Ref.~\onlinecite{flipse_direct_2012}.

First, a pair of 40-nm-thick Pt Joule heaters, which are 400 nm apart, were deposited. Then an 8-nm-thick AlO$_{x}$ layer was deposited over the sides and surfaces of the Pt Joule heaters followed by the deposition of the bottom contact (60-nm-thick Pt). The AlO$_{x}$ barriers electrically isolate the bottom contact from the Pt-heaters to avoid charge related effects. Then, the nanopillar spin valve with a structure F(15)/Cu(15)/F(15)/Au(10), where F$=$Ni$_{80}$Fe$_{20}$ or Co and the number between the parentheses are the thicknesses in nanometers, was deposited without breaking the vacuum of the deposition chamber to obtain clean interfaces. In the next two EBL steps, a top contact hole was defined followed by crosslinking a polymethyl methacrylate (PMMA) matrix around the nanopillar to isolate the bottom contact from the top contact. Finally, the top contact (130-nm-thick Au) was deposited. 

The measurements presented in this paper are all performed at room temperature using standard lock-in techniques. A low frequency ($f$=17 Hz) ac-current I$=$I$_{0}$sin$(2\pi ft)$ was used for the measurements to allow for efficient thermalization and a steady state condition. To fully characterize the samples, two different measurements were performed. First, in the spin valve measurements, the four-probe resistance of the nanopillar was measured as a function of magnetic field. To that end, a 0.1 mA current was sent through the nanopillar from contact 3 to 4 while the voltage is measured using contacts 1 and 2. From the spin valve signal, the bulk spin polarization $P_{\sigma}$, which is later used in the determination of $P_{S}$, was extracted. In thermal spin injection measurements, the open-circuit voltage across the nanopillar was measured using contacts 1 and 2 while a current of 1 mA was sent through the Pt Joule heaters (contacts 5-6 and 7-8). The measured voltage was fed to two different lock-in amplifiers which were set to record the first harmonic $V^{(1f)}$$\propto$ $I$ and second harmonic $V^{(2f)}$$\propto$ $I^{2}$ responses of the signal. In the spin valve measurement, we looked at the $V^{(1f)}$ voltage while in the thermal spin injection measurements we were mainly interested in the $V^{(2f)}$ component of the measured voltage since the spin-dependent Seebeck effect scales with the temperature gradient $\nabla T \propto I^{2}$.\cite{slachter_thermally_2010,bakker_interplay_2010,flipse_direct_2012}

\section{\label{sec:level4}Results and discussion}
\begin{figure}
\includegraphics[height=7.6cm]{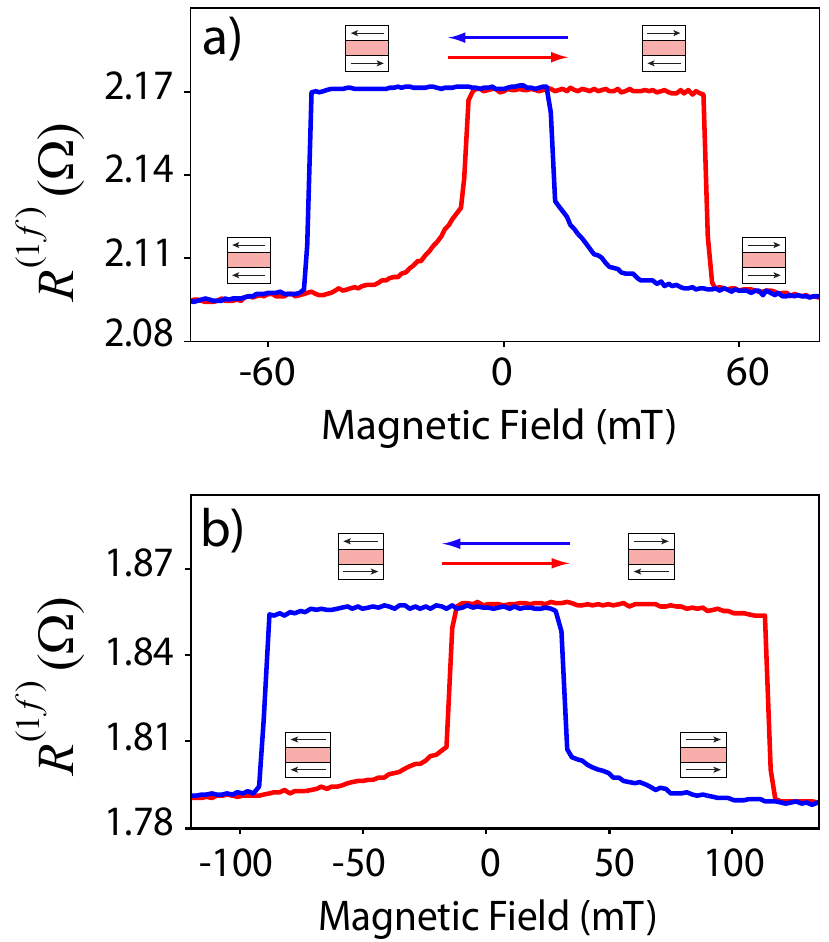}
\caption{\label{fig:fig4}(Color online) spin valve resistance $V^{(1f)}/I$ for (a) Ni$_{80}$Fe$_{20}$ and (b) Co at a current of 0.1 mA. Magnetostatic or dipolar coupling between the two magnetizations in the nanopillar favors the AP configuration at zero magnetic field.}
\end{figure}
Fig.~\ref{fig:fig4} shows the four-probe resistance $R^{(1f)}$$=$$V^{(1f)}/I$ measurements for Ni$_{80}$Fe$_{20}$ and Co nanopillar pillar devices as a function of the in-plane 
magnetic field. The spin valve signal is defined as $R_{s}^{(1f)}$=$R^{\uparrow \uparrow}$$-$$R^{\uparrow \downarrow}$, where $R^{\uparrow \uparrow}$ and $R^{\uparrow \downarrow}$ are the resistance of the pillar in the parallel and antiparallel configurations, respectively. For Ni$_{80}$Fe$_{20}$ (Fig.~\ref{fig:fig4}(a)), a spin valve signal of $-$75 m$\Omega$ was observed on top of a background resistance, $R_{b}^{(1f)}$$=$$(R^{\uparrow \uparrow}$+$R^{\uparrow \downarrow}$)/2, of 2.13 $\Omega$. By using the measured spin signal as the only fitting parameter in the 3D-FEM, a conductivity polarization $P_{\sigma}$ of 0.46 was extracted, which is in agreement with Andreev reflection measurements.\cite{soulen_andreev_1999} The calculated background resistance $R_{b}^{(1f)}$ of 1.77 $\Omega$ calculated with the finite element model is in reasonable agreement with the measured background resistance.

The input parameters to the finite element model, which are $\sigma$, S, $\kappa$ and $\Pi$, were all know from measurements in separate dedicated devices. The spin relaxation lengths $\lambda_{F}$ for Ni$_{80}$Fe$_{20}$ and Co were obtained from Ref.~\onlinecite{[][{, and references therein.}]bass_spin-diffusion_2007}. We used a spin relaxation length $\lambda_F$ of 5 nm for Ni$_{80}$Fe$_{20}$ and 40 nm for Co, respectively. These values were systematically chosen by calculating the spin signal for different values of spin relaxation lengths and fitting it to the measured spin signals (See Fig.~\ref{fig:fig8} in Sec.~\ref{appendix3}).

Following similar analysis procedure for Co (Fig.~\ref{fig:fig4}(b)), from a spin signal $R_{s}^{(1f)}$ of $-$60 m$\Omega$, we found a conductivity polarization $P_{\sigma}$$=$0.45 in agreement with Andreev reflection measurements in metallic point contacts\cite{soulen_andreev_1999} and values reported elsewhere\cite{gravier_thermodynamic_2006}. The background resistance, $R_{b}^{(1f)}$$=$1.82 $\Omega$, obtained from the measurement is a factor of two higher than the calculated background resistance of 0.99 $\Omega$. This points to the presence of a possible interfacial resistance at the bottom Pt/Co or top Co/Au interfaces, which can effectively increase the resistance of the stack. Such resistive layer may arise, for example, from interfacial disorder due to some lattice mismatch, atomic or magnetic disorders.\cite{Tsymbal_Handbook_2011} If we account for such interfacial resistance, for a conductivity polarization $P_{\sigma}$$=$0.52, we obtain a background resistance $R_{b}^{(1f)}$ of 1.5 $\Omega$ and a spin valve signal $R_{s}^{(1f)}$ of $-$56 m$\Omega$ in good agreement with the measurement.
\begin{figure}
\includegraphics[height=7.6cm]{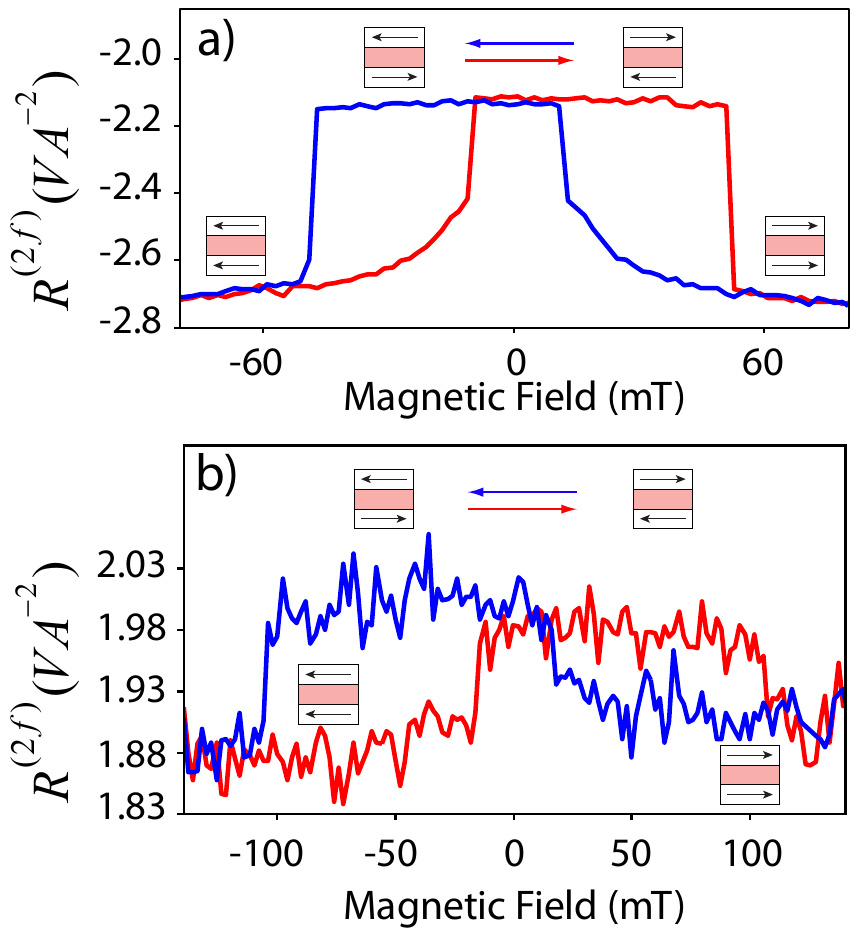}
\caption{\label{fig:fig5}(Color online) Spin-dependent Seebeck resistance $V^{(2f)}/I^{2}$ for (a) Ni$_{80}$Fe$_{20}$ and (b) Co at a current of 1 mA. Clear jumps in the measured voltage across the nanopilalr occur at fields where the two magnetizations switch.}
\end{figure}

Fig.~\ref{fig:fig5} shows the spin-dependent Seebeck measurements for a charge current of 1 mA through each Pt Joule heaters (contacts 5 to 6 and 7 to 8) in opposite directions. The heat generated from the dissipated power in the Pt Joule heaters diffuses through the AlO$_{x}$ insulating barrier and heats the bottom of the nanopillar thereby creating a temperature gradient over the stack. The temperature gradient across the pillar creates a Seebeck voltage $V^{(2f)}$ that depends on the relative orientation of the two magnetizations in the nanopillar. 
\begin{table*}
\caption{\label{TableI}Results of measurement on six other samples. The measured spin signals $R_{s}^{(1f)}$, $R_{s}^{(2f)}$, and background resistances $R_{b}^{(1f)}$, $R_{b}^{(2f)}$ are presented together with th calculated $R_{b, calc}^{(1f)}$ and $R_{b, calc}^{(2f)}$ (shaded columns). The extracted polarization of the conductivity $P_{\sigma}$ and the Seebeck coefficient $P_{S}$ are also shown.}
\begin{ruledtabular}
\begin{tabular}{lccc>{\columncolor[gray]{0.8}[\tabcolsep]}cc>{\columncolor[gray]{0.8}[\tabcolsep]}ccclr}
Sample& $R_{s}^{(1f)}$ & $R_{s}^{(2f)}$ & $R_{b}^{(1f)}$ & $R_{b, calc}^{(1f)}$ & $R_{b}^{(2f)}$ & $R_{b, calc}^{(2f)}$& $P_{\sigma}$$=$$\frac{\sigma_{\uparrow}-\sigma_{\downarrow}}{\sigma_{F}}$& $P_{S}$$=$$\frac{S_{\uparrow}-S_{\downarrow}}{S_{F}}$&$S_{\uparrow}$$-$$S_{\downarrow}$\\
 & \footnotesize(m$\Omega$) & \footnotesize(V A$^{-2}$) & \footnotesize($\Omega$) & \footnotesize($\Omega$) & \footnotesize(V A$^{-2}$) & \footnotesize(V A$^{-2}$)& & &\footnotesize($\mu$V K$^{-1}$) \\
\hline
Py (Presented in main text)&-75&-0.60&2.12&1.77&-2.4&-2.43&0.46&0.25&-4.50\\
Py1&-61&-0.70&1.85&1.76&-4.0&-2.48&0.42&0.26&-4.68\\
Py2&-70&-0.60&2.26&1.76&-3.9&-2.42&0.45&0.25&-4.50\\
Py3&-80&-0.65&1.90&1.77&-4.0&-2.45&0.47&0.25&-4.50\\
Co (Presented in main text)&-60&-0.12&1.82&0.99&1.93&6.23&0.45&0.07&-1.68\\
Co1&-60&-0.12&1.89&0.99&1.64&6.23&0.45&0.07&-1.68\\
Co2&-62&-0.13&1.82&0.99&2.0&6.28&0.45&0.08&-1.92\\
Co3&-65&-0.12&1.83&1.02&1.95&6.23&0.46&0.07&-1.68\\
\end{tabular}
\end{ruledtabular}
\end{table*}

For Ni$_{80}$Fe$_{20}$ (Fig.~\ref{fig:fig5}(a)), a spin-dependent Seebeck signal $R_{s}^{(2f)}$ of $-$0.6 VA$^{-2}$ was measured on top of a background resistance $R_{b}^{(2f)}$$=$$-$2.4 V$ $A$^{-2}$. From the measured spin signal, we obtain a spin-dependent Seebeck coefficient $S_{S}$$=$$S_{\uparrow}$$-$$S_{\downarrow}$ of $-$4.5 $\mu$V K$^{-1}$ corresponding to a Seebeck coefficient polarization $P_{S}$$=$$(S_{\uparrow}-S_{\downarrow})/S_{F}$ of 0.25 in agreement with previous reports\cite{slachter_thermally_2010,flipse_direct_2012}, where $S_{F}$$=$($\sigma_{\uparrow}S_{\uparrow}$$+$$\sigma_{\downarrow}S_{\downarrow}$)/$\sigma_{F}$.\cite{slachter_thermally_2010,flipse_direct_2012} The negative sign indicates that the Seebeck coefficient of spin-up electrons, which are the majority spins in Ni$_{80}$Fe$_{20}$ and Co, is more negative than that of the spin-down electrons. The calculated background resistance $R_{b}^{(2f)}$ of $-$2.43 V A$^{-2}$ is in good agreement with the measured background resistance.

For cobalt (Fig.~\ref{fig:fig5}(b)), for a heating current of 1 mA, a spin signal $R_{s}^{(2f)}$ of $-$0.12 V A$^{-2}$ was obtained. Similar analysis gives a spin-dependent Seebeck coefficient $S_{S}$ of $-$1.7 $\mu$V K$^{-1}$ that corresponds to a Seebeck polarization $P_{S}$$=$0.07. This result is comparable with a tight-binding calculation of the Seebeck coefficient of Co/Cu multilayers\cite{tsymbal_sign_1999} where, from the energy derivative of $\sigma$ and Mott's relation for the Seebeck coefficient, a Seebeck coefficient difference of $-$1.76 $\mu$V K$^{-1}$ between the parallel and antiparallel configurations was obtained.
The measured background resistance $R_{b}^{(2f)}$ of 1.93 V A$^{-2}$ is lower than the calculated $R_{b}^{(2f)}$ of 6.23 V$ $A$^{-2}$. This discrepancy can be again attributed to the extra interfacial resistive layer that can modify the heat current (temperature profile) across the stack. Taking this interfacial thermal resistance in to account, we obtain a background resistance $R_{b}^{(2f)}$ of 2.4 V A$^{-2}$ in good agreement with the measurement. The Seebeck coefficient polarization $P_{S}$ of 0.14 obtained is however two times higher than that obtained without including the interfacial resistance ($P_{S}$$=$0.07). 

In Fig.~\ref{fig:fig5}(b), there exists a visible asymmetry in the two parallel configurations due to possible contributions from spin-orbit effects like the anomalous Nernst effect.\cite{slachter_anomalous_2011}

The results presented above were for two samples, one for Ni$_{80}$Fe$_{20}$ and one for Co, from a total of eight samples which were measured in a similar manner. Table~\ref{TableI} shows the measurement results of the remaining six samples. The polarization of the conductivity $P_{\sigma}$ and Seebeck coefficient $P_{S}$ were extracted by fitting the measured spin signals to the 3D-FEM. The modeled background resistances, which are shown in shaded columns, are in reasonably good agreement with the measurements and are consistent with the samples presented in the text.
\section{\label{sec:level5}Conclusion}
In summary, we have performed all-electrical spin-dependent Seebeck effect measurements in Ni$_{80}$Fe$_{20}$ and Co nanopillar spin valve devices. We found that the polarization of the Seebeck coefficient for Ni$_{80}$Fe$_{20}$ ($\sim$25\%) and Co ($\sim$8\%) are in agreement with earlier experimental studies in Ni$_{80}$Fe$_{20}$/Cu spin valve structures and earlier theoretical works in Co/Cu multilayers, respectively. With the method presented here, it is in principle possible to measure the polarization of the conductivity and Seebeck coefficient of any ferromagnetic metal that makes up a symmetric or asymmetric spin valve.
\begin{acknowledgments}
The authors thank B. Wolfs, M. de Roosz and J.G. Holstein for technical assistance and N. Vlietstra for reading the manuscript. This work is part of the research program of the Foundation for Fundamental Research on Matter (FOM) and is supported by NanoLab NL, EU FP7 ICT (grant no. 257159 MACALO) and the Zernike Institute for Advanced Materials.
\end{acknowledgments}
\appendix
\section{\label{appendix1}Temperature profile across F/N/F stack}
\begin{figure}
\includegraphics[width=8.6cm]{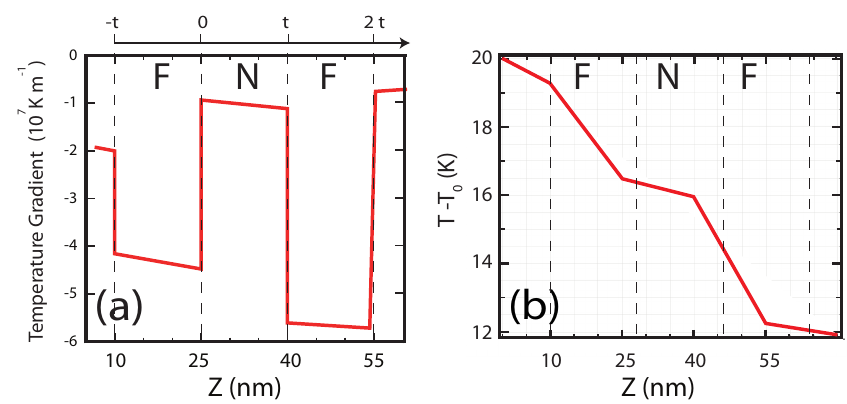}
\caption{\label{fig:fig6}(a) Temperature gradient in a F/N/F pillar spin valve stack and (b) The temperature profile across the F/N/F stack for a heating current of 2 mA through both Joule-heaters. For 1 mA current, the scale reduces by a factor of four.}
\end{figure}

Fig.~\ref{fig:fig6} shows the temperature gradient and temperature profile of a symmetric F/N/F stack.
From the FEM, for a heating current of 2 mA through the Pt Joule heaters, a temperature gradient up to 40 K $\mu$m$^{-1}$ can be achieved in our devices (see Fig.~\ref{fig:fig6}(a)) corresponding to a $\Delta T$$=$ 8 K across the F/N/F stack (see Fig.~\ref{fig:fig6}(b)). The red line in Fig.~\ref{fig:fig6}(a) shows the temperature gradient across a Ni$_{80}$Fe$_{20}$/Cu/Ni$_{80}$Fe$_{20}$ pillar spin valve. From continuity of the heat current $\vec{Q}$$=$$-\kappa\vec{\nabla} T$ at the F/N interfaces, the temperature gradient in the ferromagnetic region $\nabla T_{F}$ is related to that of the N-region $\nabla T_{N}$ as:
\begin{equation}
\vec{\nabla} T_{F}=\frac{\kappa_{N}}{\kappa_{F}}\vec{\nabla}T_{N},
\label{eq:gradientF}
\end{equation}
where $\kappa_{F}$ and $\kappa_{N}$ are the thermal conductivity of the F- and N-region, respectively.
\section{\label{appendix2}Expression for the spin accumulation}
To obtain an expression for the spin accumulation $\mu_{s}$$=$$\mu_{\uparrow}$$-$$\mu_{\downarrow}$, we first need to solve the Valet-Fert 1D-spin diffusion equation $\frac{\partial^{2}(\mu_{\uparrow}-\mu_{\downarrow})}{\partial z^{2}}$$=$$\frac{\mu_{\uparrow}-\mu_{\downarrow}}{\lambda_{sf}^{2}}$ for each region in the F/N/F stack\cite{slachter_thermally_2010}, where $\lambda_{sf}$ is the spin relaxation length. The general solutions for each region reads:

\noindent Region I: ($-t< z <0$)
\begin{equation}
\begin{aligned}
\mu_{\uparrow,\downarrow}=A+Bz\pm\frac{C}{\sigma_{\uparrow,\downarrow}}e^{-z/\lambda_{F}}\pm\frac{D}{\sigma_{\uparrow,\downarrow}}e^{z/\lambda_{F}},
\end{aligned}
\label{eq:appendixI}
\end{equation}
Region II: ($0< z <t$)
\begin{equation}
\begin{aligned}
\mu_{\uparrow,\downarrow}=Fz\pm\frac{2G}{\sigma_{N}}e^{-z/\lambda_{N}}\pm\frac{2H}{\sigma_{N}}e^{z/\lambda_{N}},
\end{aligned}
\label{eq:appendixII}
\end{equation}
Region III: ($t< z < 2t$)
\begin{equation}
\begin{aligned}
\mu_{\uparrow,\downarrow}=K+Lz\pm\frac{M}{\sigma_{\uparrow,\downarrow}}e^{-z/\lambda_{F}}\pm\frac{N}{\sigma_{\uparrow,\downarrow}}e^{z/\lambda_{F}},
\end{aligned}
\label{eq:appendixIII}
\end{equation}
where $+$ and $-$ denote the spin-up and spin-down, respectively, $\lambda_{F}$ and $\lambda_{N}$ are the spin relaxation length in the F- and N-regions.
\begin{figure}
\includegraphics[width=8.7cm]{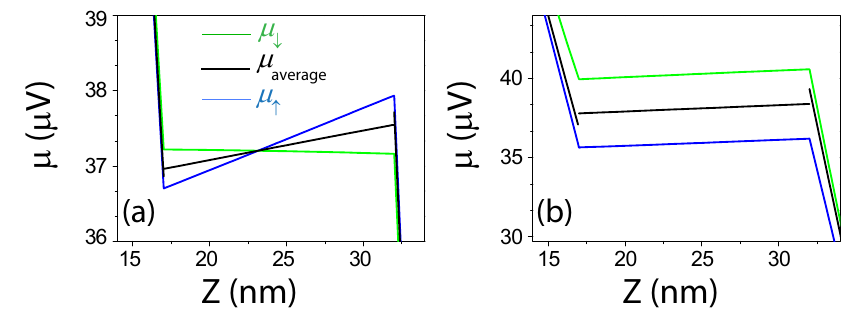}
\caption{\label{fig:fig7}(Color online). The electrochemical potential profile for spin-up $\mu_{\uparrow}$ and spin-down $\mu_{\downarrow}$ electrons and the average electrochemical potential $\mu_{\text{average}}$$=$($\mu_{\uparrow}\sigma_{\uparrow}$+$\mu_{\downarrow}\sigma_{\downarrow}$)/$\sigma_{F}$ for (a) $\uparrow\uparrow$ and (b) $\uparrow\downarrow$ configurations.}
\end{figure}
The spin accumulation $\mu_{s}$ at $z$$=$0 and $z$$=$$t$ can then be expressed as a function of these coefficients as $\mu_{s}$$(z$=$0)$$=$$\frac{4}{\sigma_{N}}(G$$+$$H)$ and $\mu_{s}$$(z$=$t)$$=$$\frac{4}{\sigma_{N}}(Ge^{-z}$$+$$He^{z})$, respectively. For a symmetric spin valve the spin accumulation, for example, at interface $z$$=$0 for the $\uparrow \uparrow$ and $\uparrow \downarrow$ configurations reads:
\begin{widetext}
\begin{eqnarray}
\mu_{s}^{\uparrow \uparrow}(z=0)=-\text{e$\lambda $}_FS_S\nabla T\frac{\left[\text{coth}(\frac{t}{\lambda _N})+\frac{\exp({-\frac{2t}{\lambda _F}})}{\text{sinh}(\frac{t}{\lambda _N})}-\text{tanh}(\frac{t}{2 \lambda _N})\text{cosh}(\frac{t}{\lambda _F})\right] \frac{R_F}{R_N}+ \left[-\text{sinh}(\frac{t}{\lambda _F})+\text{tanh}(\frac{t}{\lambda _F})\right]}{\text{cosh}(\frac{t}{\lambda _F})\left[\frac{R_F^2}{R_N^2}+2\text{coth}(\frac{t}{\lambda _N}) \text{tanh}(\frac{ t}{\lambda _F}) \frac{R_F}{R_N}+\text{tanh}(\frac{t}{\lambda _F}){}^2\right]},\\
\mu_{s}^{\uparrow \downarrow}(z=0)=-\text{e$\lambda $}_FS_S\nabla T\frac{\left[\text{coth}(\frac{t}{\lambda _N})+\frac{\exp({-\frac{2t}{\lambda _F}})}{\text{sinh}(\frac{t}{\lambda _N})}-\text{coth}(\frac{t}{2 \lambda _N})\text{cosh}(\frac{t}{\lambda _F})\right] \frac{R_F}{R_N}+ \left[-\text{sinh}(\frac{t}{\lambda _F})+\text{tanh}(\frac{t}{\lambda _F})\right]}{\text{cosh}(\frac{t}{\lambda _F})\left[\frac{R_F^2}{R_N^2}+2\text{coth}(\frac{t}{\lambda _N}) \text{tanh}(\frac{ t}{\lambda _F}) \frac{R_F}{R_N}+\text{tanh}(\frac{t}{\lambda _F}){}^2\right]}
\label{eq:appendix3}
\end{eqnarray}
\end{widetext}
In the limit $\lambda_{N},\lambda_{F}$$\ll$ $t$, Eq.~\ref{eq:appendix3} reduces to the result obtained for a single F/N interface given in Ref.~\onlinecite{slachter_thermally_2010,scharf_theory_2012}. Fig.~\ref{fig:fig7} shows the chemical potential profile across a F/N/F spin valve for the $\uparrow\uparrow$ and $\uparrow\downarrow$ configurations as obtained from the FEM. At the F/N interfaces, for both $\uparrow\uparrow$ and $\uparrow\downarrow$ configurations, a discontinuity in the average electrochemical potential $\mu_{\text{average}}$ leads to an electrochemical potential drop $\Delta \mu$$=$$P_{\sigma} \mu_{s}$. The spin valve signal $V_{SV}$ is expressed in terms of these electrochemical potential drops as:
\begin{equation}
V_{SV}=((\Delta \mu_{z=0}^{\uparrow \uparrow}+\Delta \mu_{z=t}^{\uparrow \uparrow})-(\Delta \mu_{z=0}^{\uparrow \downarrow}+\Delta \mu_{z=t}^{\uparrow \downarrow}))/e
\label{eq:appendix4}
\end{equation}
\section{\label{appendix3}Material parameters used in 3D-FEM}
\begin{figure}[h]
\includegraphics[width=8.6cm]{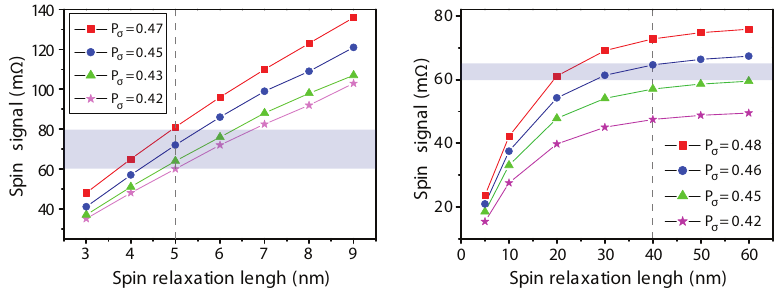}
\caption{\label{fig:fig8}(Color online) Dependence of the spin valve signal on the spin relaxation length $\lambda_F$ of the FM for (a) Ni$_{80}$Fe$_{20}$/Cu/Ni$_{80}$Fe$_{20}$ and (b) Co/Cu/Co nanopillar spin valves. A $\lambda_F$ of 5 nm for Ni$_{80}$Fe$_{20}$ and 40 nm for Co fits the measured spin signal, shown by the shaded region.}
\end{figure}

One important aspect of the finite-element modeling is good knowledge of the temperature and 
\begin{table}[h]
\caption{\label{TableII}Material parameters used in finite element modeling. The spin relaxation length $\lambda_{s}$ was taken from various sources of literature.\cite{dubois_evidence_1999, bass_spin-diffusion_2007}} 
\begin{ruledtabular}
\begin{tabular}{cccccc}
Material&$t$& $\sigma$ & $\kappa$ & $S$ & $\lambda_{s}$ \\
 &(nm)& ($10^{6} $ S m$^{-1}$) & (Wm$^{-1}$K$^{-1}$) & ($\mu$VK$^{-1}$) & (nm) \\
\hline
Ni$_{80}$Fe$_{20}$ &15 & 2.9 & 17 & -18 & 5 \\
Co &15 & 6.0 & 40 & -22 & 40 \\
Cu &15 & 15 & 10 & 1.6 & 300  \\
Pt &40 & 4.2 & 32 & -5 & 5 \\
Pt &60 & 4.8 & 37 & -5 & 3 \\
Au &120 & 27 & 180 & 1.7 & 80 \\
AlO$_{x}$ &8& 10$^{-18}$ & 0.12 & 0 & - \\
SiO$_{2}$ &300& 10$^{-19}$ & 1 & 0 & - 
\end{tabular}
\end{ruledtabular}
\end{table}
voltage profiles in the F/N/F pillar devices. This requires usage of appropriate material parameters in the 3D-FEM, which can often lead to underestimating background electrical and thermal voltages if bulk material parameters were used.\cite{slachter_thermally_2010,bakker_interplay_2010}
 Table~\ref{TableII} shows  material parameters used in the model. Electrical conductivity of each material was measured using a standard four probe geometry. The thermal conductivity was then calculated using the Wiedemann-Franz law. For device dimensions discussed in the main text, the electronic contribution to the thermal conductivity is dominant over the lattice (phononic) conductivity\cite{bakker_thermal_2012}. The Seebeck coefficients were measured by using the technique presented in Ref.~\onlinecite{bakker_thermal_2012}. One parameter which was not measured but obtained from literature is the spin relaxation length $\lambda_{F}$ of the ferromagnets. The spin relaxation length for Ni$_{80}$Fe$_{20}$ of 5 nm is well established in literature.~\cite{dubois_evidence_1999,[][{ and references therein.}]bass_spin-diffusion_2007} However, reported spin relaxation length of Co at room temperature vary from 20 nm to 60 nm\cite{dubois_evidence_1999,bass_spin-diffusion_2007}. The spin valve signals that are extracted from the model depend on the spin relaxation length and the polarization of the conductivity. To tackle the uncertainty in the spin relaxation length in Co, we performed a calculation of the spin signal for varying spin relaxation length values of the ferromagnet. Fig.~\ref{fig:fig8} shows the dependence of the spin signal on the spin relaxation length for different values of the conductivity polarization $P_{\sigma}$ ranging between 0.42 and 0.47 (for Ni$_{80}$Fe$_{20}$) and 0.42 and 0.48 (for Co). 
The shaded region in the figures indicates the region in which the measured spin signal values fall.
For a choice of spin relaxation lengths of 5 nm (for Ni$_{80}$Fe$_{20}$) and 40 nm (for Co), the measured spin valve signals can be well fitted with the model. Hence, we used these two values for the determination of the spin-dependent Seebeck coefficients.
\bibliography{References}
\end{document}